\documentclass[twocolumn]{aastex701}

\usepackage{natbib}
\usepackage{url}
\usepackage{threeparttable}
\usepackage{comment}
\renewcommand{\arraystretch}{1.3} 
\usepackage{booktabs}
\usepackage{makecell}
\graphicspath{{./}}
\usepackage{multirow} 

\begin{document}

\title{Tracing the Orbital Motion of the Accreting White Dwarf in EX~Hydrae with XRISM/Resolve}

\correspondingauthor{Yuken Ohshiro}

\author[orcid=0009-0002-4783-3395,gname=Yuken,sname=Ohshiro]{Yuken Ohshiro}
\affiliation{RIKEN Pioneering Research Institute (PRI), 2-1 Hirosawa, Wako, Saitama 351-0198, Japan}
\email[show]{ohshiro.yuken@gmail.com}

\author[gname=Yukikatsu,sname=Terada]{Yukikatsu Terada}
\affiliation{Department of Physics, Graduate School of Science and Engineering, Saitama University, Saitama, 338-8570, Japan}
\affiliation{Institute of Space and Astronautical Science (ISAS), Japan Aerospace Exploration Agency (JAXA), 3-1-1 Yoshinodai, Chuo-ku, Sagamihara, Kanagawa 252-5210, Japan}
\email{terada@mail.saitama-u.ac.jp}

\author[gname=Taichi,sname=Ichikawa]{Taichi Ichikawa}
\affiliation{Department of Physics, Graduate School of Science and Engineering, Saitama University, Saitama, 338-8570, Japan}
\email{taichi.ichikawa@saitama-u.ac.jp}

\author[gname=Yugo,sname=Motogami]{Yugo Motogami}
\affiliation{Department of Physics, Graduate School of Science and Engineering, Saitama University, Saitama, 338-8570, Japan}
\email{y.motogami.738@ms.saitama-u.ac.jp}

\author[gname=Manabu,sname=Ishida]{Manabu Ishida}
\affiliation{Institute of Space and Astronautical Science (ISAS), Japan Aerospace Exploration Agency (JAXA), 3-1-1 Yoshinodai, Chuo-ku, Sagamihara, Kanagawa 252-5210, Japan}
\email{ishidamnb@gmail.com}

\author[gname=Koji,sname=Mukai]{Koji Mukai}
\affiliation{CRESST and X-ray Astrophysics Laboratory, NASA/GSFC, Greenbelt, MD 20771, USA}
\affiliation{Department of Physics, University of Maryland, Baltimore County, 1000 Hilltop Circle, Baltimore, MD 21250, USA}
\email{koji.mukai@nasa.gov}

\author[gname=Masayoshi,sname=Nobukawa]{Masayoshi Nobukawa}
\affiliation{Faculty of Education, Nara University of Education, Nara, 630-8502, Japan}
\email{nobukawa@cc.nara-edu.ac.jp}

\author[gname=Takayuki,sname=Hayashi]{Takayuki Hayashi}
\affiliation{Department of Physics, Kyoto University, Kyoto, Japan}
\email{hayashi.takayuki.3f@kyoto-u.ac.jp}

\author[gname=Mariko,sname=Kimura]{Mariko Kimura}
\affiliation{Advanced Research Center for Space Science and Technology, Kanazawa University, Ishikawa, Japan}
\email{mariko-kimura@se.kanazawa-u.ac.jp}

\author[gname=Mai,sname=Takeo]{Mai Takeo}
\affiliation{Faculty of Science, Graduate School of Science and Engineering,University of Toyama, Toyama, Japan}
\email{takeo@sci.u-toyama.ac.jp}

\begin{abstract}
    Measuring the masses of accreting white dwarfs (WDs) is crucial for understanding their evolution and the physics of accretion.
    High-resolution X-ray spectroscopy can trace the WD motion through Doppler shifts of emission lines formed close to the WD.
    We report an 83~ks XRISM/Resolve observation of the intermediate polar EX~Hydrae and measure the orbital modulation of individual Fe K-shell line centroids.
    The Fe~{\sc xxv} K$\alpha$ components show coherent orbital modulation, yielding $K_1 = 58.1 \pm 8.5\ \mathrm{km\ s^{-1}}$.
    This is the first detection of orbital modulation in individual Fe K-shell lines from an accreting WD, made possible by the high spectral resolution of Resolve and its frequent in-orbit gain calibration.
    The measured $K_1$ is consistent with optical/UV $K_1$ measurements, providing a cross-check that these distinct tracers follow the WD orbital motion.
    Combining this X-ray measurement with literature orbital parameters, we derive a WD mass of $M_1 = 0.79 \pm 0.04\ M_\odot$.
    These results demonstrate that high-resolution X-ray spectroscopy can use individual Fe K-shell line centroids to trace WD orbital motion in accreting WDs.
\end{abstract}


\section{Introduction} \label{sec: intro}

Accreting white dwarfs (WDs) in binary systems, including cataclysmic variables (CVs), are important laboratories for studying accretion physics and binary evolution \citep[e.g.,][]{mukai2017,pala2022}.
Reliable WD mass measurements also connect individual systems to broader questions such as Type Ia supernova progenitor channels \citep[e.g.,][]{ruiter2025} and the origin of the Galactic ridge X-ray emission \citep[e.g.,][]{revnivtsev2009,mukai2017}.

Dynamical measurements provide the most direct route to the masses of accreting WDs because they constrain the orbital motion of the binary components \citep[e.g.,][]{thoroughgood2004,savoury2011}.
The radial-velocity semi-amplitude of the accreting WD, $K_1$, is a key quantity in the binary mass function.
Optical and ultraviolet spectroscopy can provide powerful dynamical constraints when the donor star, WD photosphere, or eclipses constrain the binary geometry.
When $K_1$ is inferred from accretion-related emission lines, however, the central issue is to separate the WD orbital motion from line components formed in the disk, stream or bright spot, irradiated material, and, in magnetic systems, magnetically controlled flow \citep[e.g.,][]{marsh1988,north2002,rodriguez-gil2007}.

High-resolution X-ray line spectroscopy offers a complementary way to probe the WD motion by observing line-emitting plasma much closer to the accreting WD.
In magnetic cataclysmic variables, and in particular in intermediate polars, the WD magnetosphere channels the inner accretion flow onto the magnetic poles, where a stand-off shock forms close to the WD surface \citep[e.g.,][]{aizu1973,mukai2017}.
The shock heats the accreting gas to X-ray-emitting temperatures, producing strong Fe~{\sc xxv} K$\alpha$ and Fe~{\sc xxvi} Ly$\alpha$ lines in the post-shock plasma \citep[e.g.,][]{yuasa2010,hayashi2014}, while Fe K$\alpha$ fluorescence is produced by irradiation of nearby low-ionization material, including the WD surface \citep[e.g.,][]{hayashi2018}.
Because these Fe K-shell lines arise in the immediate accretion environment of the WD, their orbital Doppler shifts can constrain $K_1$ with less sensitivity to the geometry of the extended optical/UV line-emitting flow.

EX~Hydrae provides a stringent benchmark for this X-ray dynamical test.
It is a bright, eclipsing intermediate polar with a partial X-ray eclipse \citep[e.g.,][]{rosen1988} and a line-rich X-ray spectrum suitable for high-resolution studies of post-shock plasma diagnostics \citep[e.g.,][]{luna2015}.
Its 98~min orbit, high inclination of about $78^\circ$, and WD mass of $\sim 0.78$--$0.79~M_\odot$ are well constrained by multiwavelength timing and dynamical studies \citep{beuermann2008,echevarria2016}.
Optical and ultraviolet studies have yielded consistent WD radial-velocity amplitudes near 58~km~s$^{-1}$ \citep{belle2003,echevarria2016,beuermann2024}.
Pioneering Chandra/HETG work established the feasibility of X-ray radial-velocity measurements by constructing an average line profile from many emission lines in velocity space \citep{hoogerwerf2004}.
That study measured $K_1 = 58.2 \pm 3.7$~km~s$^{-1}$, consistent with optical/UV measurements, but the signal was obtained from a many-line average in four broad orbital-phase bins rather than from individual emission lines.

The Resolve microcalorimeter onboard XRISM now enables this test with individual Fe K-shell lines.
Resolve combines high spectral resolution in the Fe K band \citep{porter2025} with in-orbit gain calibration suitable for precise line-centroid measurements \citep{eckart2025}.
Its larger Fe K-band effective area than Chandra/HETG enables modulation searches in individual Fe K-shell lines without relying on a many-line average.
Resolve has also demonstrated precise Fe K-shell diagnostics in the magnetic CV AM~Her \citep{terada2026} and orbitally modulated line-shift measurements in the X-ray binaries Cen~X-3 and Cyg~X-3 \citep{mochizuki2024,miura2025}.
Here we use XRISM/Resolve to measure the orbital modulation of component-resolved Fe K-shell line centroids in EX~Hya and test whether the resulting X-ray velocity amplitude traces the WD orbital motion.
This provides a complementary X-ray constraint on the WD radial-velocity amplitude, $K_1$, and a cross-check on optical/UV dynamical measurements.

\section{Observation and Data Reduction} \label{sec: method}

The XRISM observation (PI: Taichi Ichikawa; Observation ID: 201058010) of EX~Hydrae was performed from 18:25 UT on 2025 July 12 to 17:16 UT on 2025 July 14, with a pointing position at $(\mathrm{R.A.,~Dec.}) = (193.10133,~-29.24892)$ and a roll angle of $294^\circ$.
We analyze data from Resolve, the microcalorimeter onboard XRISM \citep{ishisaki2025,kelley2025}.
The $6\times 6$ Resolve detector array, placed at the focus of the Resolve X-ray Mirror Assembly \citep{hayashi2024}, covers a $3' \times 3'$ field of view and provides the high spectral resolution needed to measure orbital modulation of X-ray emission-line centroids \citep{porter2025}.
During the observation, Resolve was operated without an additional filter, and the gate valve with a $\sim$\,250\,$\mu$m-thick beryllium window \citep{midooka2021} was closed, limiting the bandpass to energies above $\sim$\,1.6\,keV.

The data were reprocessed by the automated XRISM processing pipeline version 03.02.014.010.
In the following data reduction, we used \texttt{HEASoft} version 6.36 and CALDB version 12.
The data were reduced following the standard Resolve event-screening procedures described by \citet{mochizuki2024a}.
We extract only Grade Hp (High-resolution primary; ITYPE=0) events, excluding pixels 7 and 27 because of the anomalous gain behavior noted in the Gain Recovery Report\footnote{\url{https://heasarc.gsfc.nasa.gov/FTP/xrism/postlaunch/gainreports/2/201058010_resolve_energy_scale_report.pdf}}.
The resulting effective exposure time after event screening is 83~ks.
The gain and energy calibration of the Resolve detector is performed every orbit using the Mn K$\alpha$ line at $\simeq 5.9$~keV emitted by the onboard $^{55}$Fe calibration source; for high-resolution events, the energy-scale accuracy is $<0.3$~eV over the 5.4--9.0~keV band \citep{eckart2025}.
For this observation, the Gain Recovery Report gives an energy-scale residual below 0.1~eV for the Mn K$\alpha$ calibration line in the retained Hp events, corresponding to $\sim 4$--5~km~s$^{-1}$ near the Fe K band, well below the orbital velocity amplitude measured below.

\begin{figure*}[t]
    \centering
    \includegraphics[width=1.0\linewidth]{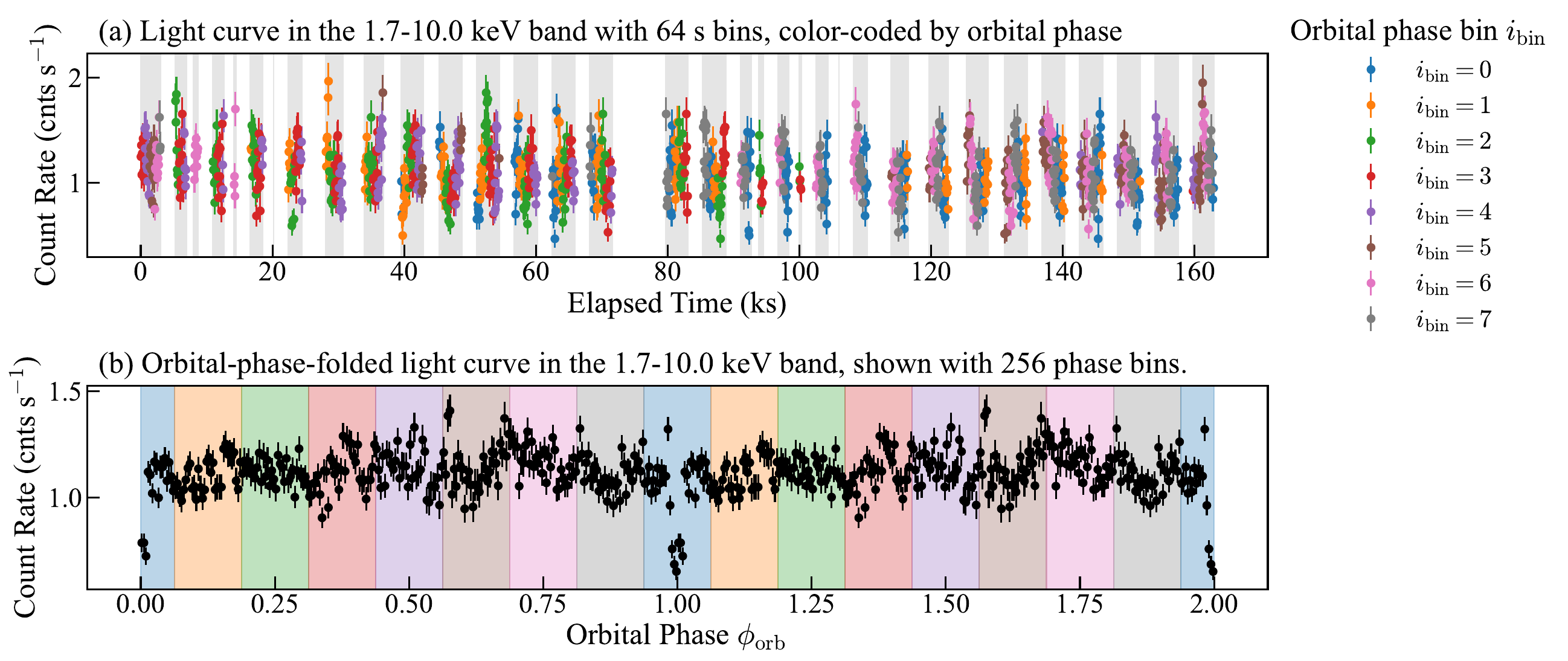}
    \vspace{-4pt}
    \caption
    {
        (a) Light curves in the 1.7--10~keV band with 64~s bins, color-coded by orbital phase for clarity. Gray shaded regions mark the good-time intervals (GTIs); only 64~s bins fully contained within GTIs are plotted. The orbital phase is defined using the BJD(TDB) eclipse ephemeris described in the text.
        (b) Folded light curve in the 1.7--10~keV band of EX~Hya with 256 phase bins per cycle.
        The data are repeated for two cycles for better visualization, and the partial X-ray eclipse appears near $\phi_{\mathrm{orb}} = 0$.
    }
    \label{fig:lightcurve}
\end{figure*}

Figure~\ref{fig:lightcurve}a shows the 1.7--10~keV light curves with 64~s time bins.
Apart from the orbital modulation, the light curves show no flares or large changes in the count rate during the observation.
We define the orbital phase using the eclipse ephemeris of \citet{echevarria2016}.
Because the original ephemeris is given in HJD without an explicitly specified time standard, we assume TT and convert the reference epoch to BJD(TDB) following \citet{eastman2010}, adopting the coordinates of EX~Hya.
The resulting ephemeris is
\[
\mathrm{BJD}_{\mathrm{eclipse}}(\mathrm{TDB}) = 2437699.941313 + 0.068233843 \times E,
\]
where $E$ is the cycle number.
The statistical uncertainty of the original ephemeris and the precision of the HJD-to-BJD conversion are both of order 1~s, sufficiently small for the present orbital-modulation analysis.

For the phase-resolved spectral analysis, we divide the orbital phase into bins defined as
\[
\phi_{\mathrm{orb}} \in \left[\frac{i_{\mathrm{bin}}-0.5}{N_{\mathrm{bin}}},\ \frac{i_{\mathrm{bin}}+0.5}{N_{\mathrm{bin}}}\right)
\]
for $i_{\mathrm{bin}} = 0, 1, \ldots, N_{\mathrm{bin}}-1$, where $N_{\mathrm{bin}}$ is the number of bins per cycle.
Figure~\ref{fig:lightcurve}b shows the folded 1.7--10~keV light curve of EX~Hya with $N_{\mathrm{bin}} = 256$ phase bins.
The partial X-ray eclipse appears at $\phi_{\mathrm{orb}} \sim 0$, confirming the adopted phase assignment.
An alternative conversion assuming HJD(UTC) gives a slight eclipse-phase offset but does not significantly affect the results; we therefore use the HJD(TT)-based ephemeris below.
We also checked spacecraft-velocity and spin-phase sampling as possible phase-dependent systematics; these effects are not expected to account for the measured modulation, as described in Appendix~\ref{app:phase_systematics}.
For the phase-resolved spectra, we generated redistribution matrix files (RMFs) with \texttt{rslmkrmf} and auxiliary response files (ARFs) with \texttt{xaarfgen} for each phase bin, assuming EX~Hya to be a point source at the source position.

In the spectral analysis, we use the \texttt{XSPEC} software version 12.15.1 \citep{arnaud1996}.
We used the solar abundance table of \citet{wilms2000} and AtomDB version 3.1.3 \citep{smith2001,foster2012}.
Spectral fitting is performed for unbinned spectra using the Cash statistic \citep{cash1979} for parameter estimation, and the errors of the best-fit parameters are calculated at the 68\% confidence level.

\section{Analysis and Results} \label{sec: results}

\begin{figure*}[t]
    \centering
    \includegraphics[width=1.0\linewidth]{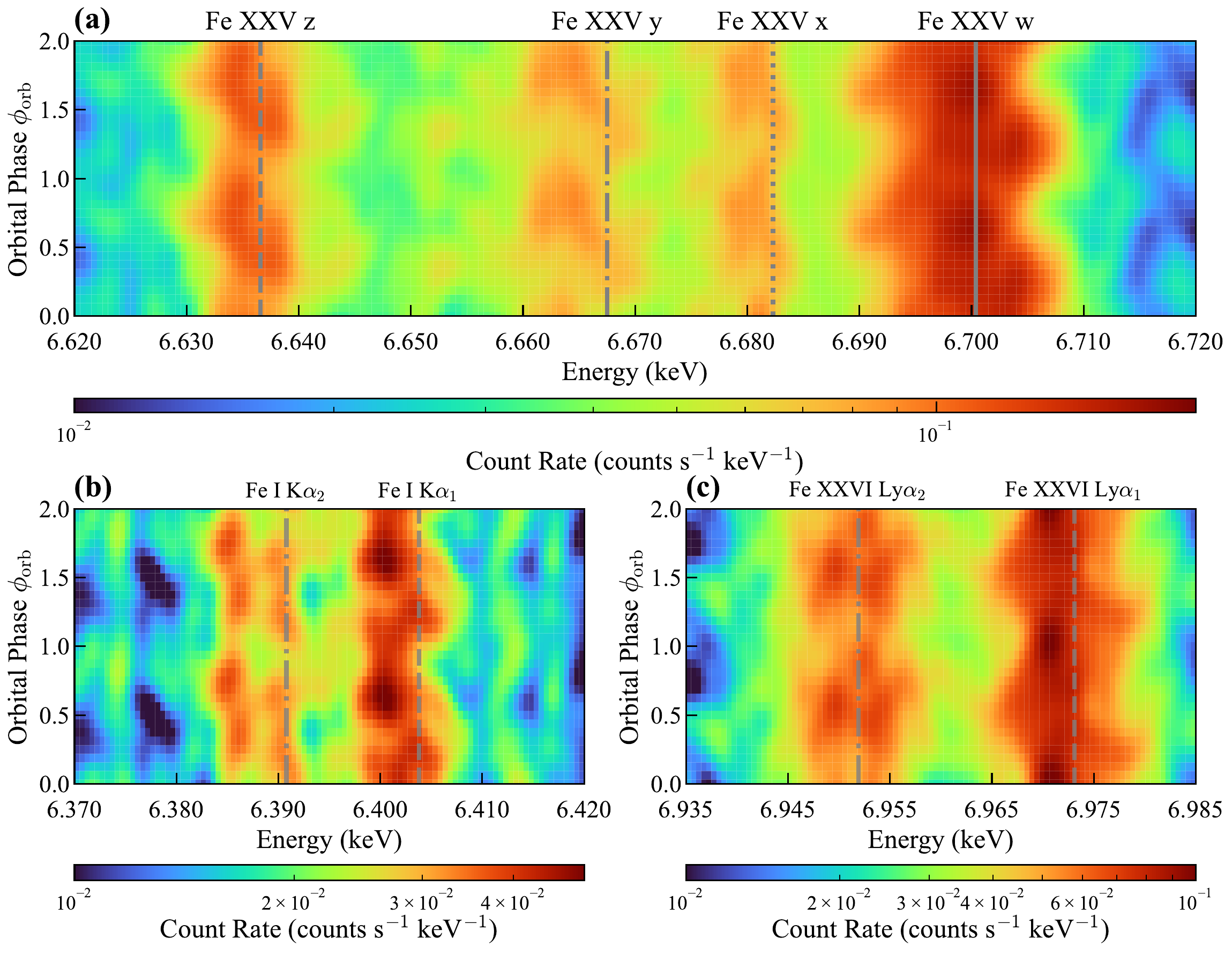}
    \vspace{-4pt}
    \caption
    {
        Count-rate maps in the Fe K band as a function of energy and orbital phase.
        The maps are constructed with an energy bin size of 0.5~eV and an orbital-phase bin size of 0.0625 (16 bins per cycle), and are displayed after Gaussian smoothing for visual clarity, with $\sigma = 2$ bins in both directions, corresponding to 1.0~eV in energy and 0.125 in orbital phase.
        The orbital phase is shown over two cycles on the vertical axis, and the color scale represents the count rate in $\mathrm{counts\ s^{-1}\ keV^{-1}}$.
        Panel (a) shows the Fe~{\sc xxv} w, x, y, and z lines, panel (b) shows the Fe K$\alpha$ fluorescence complex, and panel (c) shows the Fe~{\sc xxvi} Ly$\alpha_1$ and Ly$\alpha_2$ lines.
        The vertical gray lines indicate the rest-frame energies of the labeled transitions \citep[][]{holzer1997,yerokhin2015,yerokhin2019}.
    }
    \label{fig:feline_2dmap}
\end{figure*}

Figure~\ref{fig:feline_2dmap} provides a visual overview of the Fe K-band emission-line behavior as a function of orbital phase.
The clearest phase-dependent centroid displacement is seen in the Fe~{\sc xxv} K-shell complex, which appears most blueshifted near $\phi_{\mathrm{orb}} \simeq 0.25$ and most redshifted near $\phi_{\mathrm{orb}} \simeq 0.75$.
The Fe K$\alpha$ fluorescence complex and Fe~{\sc xxvi} Ly$\alpha_1$ show similar trends, whereas Fe~{\sc xxvi} Ly$\alpha_2$ is less visually diagnostic in the smoothed map.
This phasing is that expected for emission tied to the orbital motion of the accreting WD, motivating the phase-resolved spectral fitting below.

To quantify the orbital modulation of the line energies, we fit the orbital-phase-resolved spectra independently for each phase bin, using a bin size of $N_\mathrm{bin} = 8$ per cycle in the 6.2--7.2~keV energy band.
This band includes the Fe K$\alpha$ fluorescence complex, Fe~{\sc xxv} w, x, y, and z, and Fe~{\sc xxvi} Ly$\alpha_1$ and Ly$\alpha_2$ lines.
We model the highly ionized Fe emission with \texttt{bvcempow}, an optically thin multi-temperature plasma model with a power-law emission-measure distribution that has been applied to recent high-resolution X-ray spectroscopy of the magnetic CV AM~Her \citep{terada2026}.
We fix the emission-measure index to $\alpha = 0.5$ as a fiducial isobaric cooling-flow-like value, the density parameter to $n_{\mathrm{H}} = 1~\mathrm{cm}^{-3}$, and all elemental abundances to solar values, while allowing $kT_\mathrm{max}$, the redshift, the velocity broadening $\sigma_v$, and the normalization to vary.
The Fe K$\alpha$ fluorescence complex is modeled with \texttt{zfeklor}, which assumes a seven-Lorentzian laboratory profile of Fe~{\sc i} K$\alpha$ with fixed relative energies and intensities \citep{holzer1997}; its redshift and normalization are left free.
We do not include additional absorption or reflection components because previous broadband studies indicate that they do not dominate the Fe K-band continuum of EX~Hya \citep[e.g.,][]{luna2018}.
The NXB is negligible compared with the source count rate and does not affect the centroid-modulation analysis.

\begin{figure*}[tbp]
    \centering
    \includegraphics[width=1.0\linewidth]{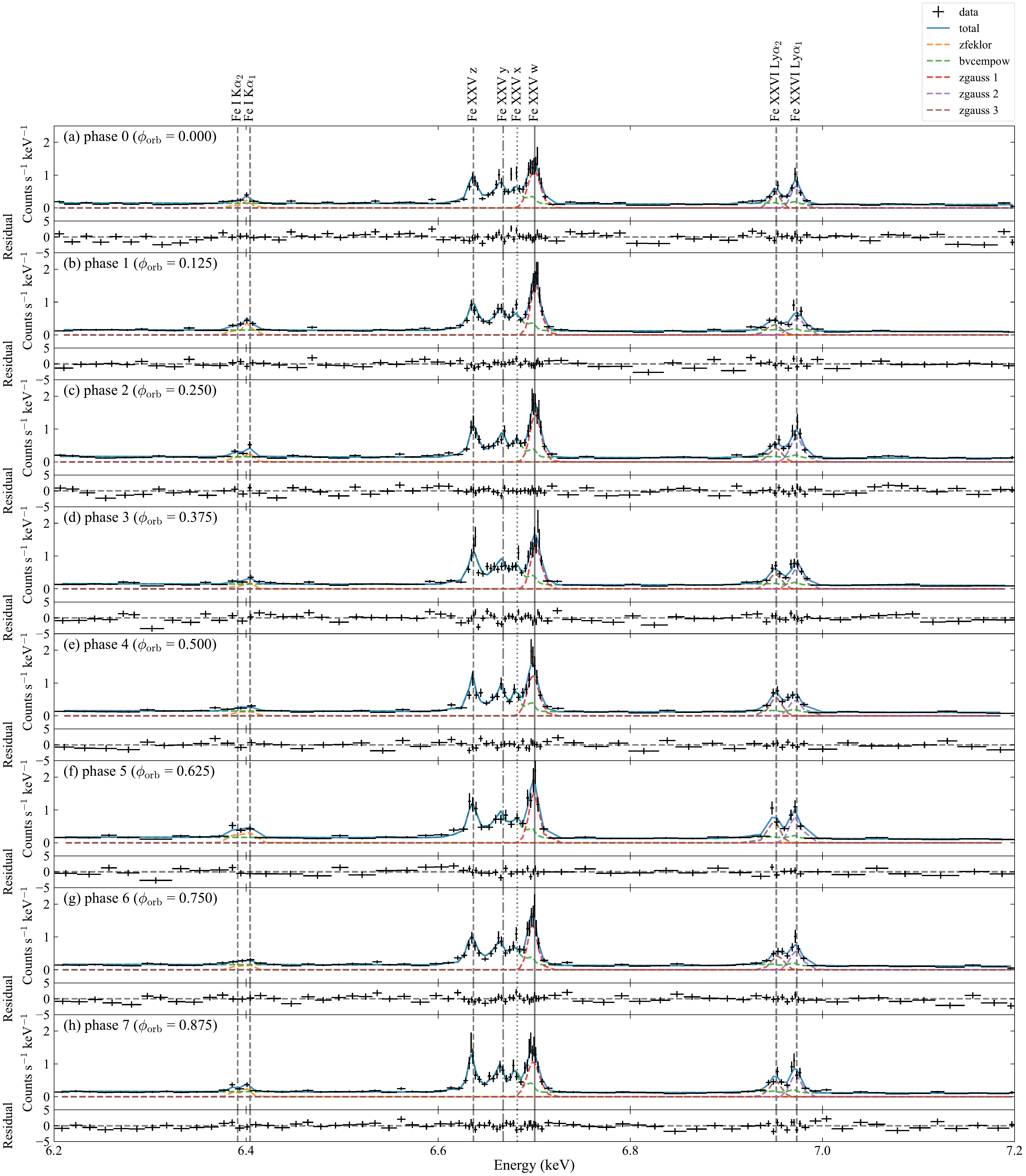}
    \vspace{-4pt}
    \caption
    {
        Phase-resolved spectra in the Fe K band (6.2--7.2~keV) with the best-fit models.
        Residuals are shown as ($\texttt{data}-\texttt{model})/\texttt{error}$.
        The vertical gray lines indicate the rest-frame energies of the labeled transitions \citep[][]{holzer1997,yerokhin2015,yerokhin2019}.
    }
    \label{fig:phase_resolved_spectra}
\end{figure*}

We first applied the model described above, in which the highly ionized Fe lines are represented by the \texttt{bvcempow} component.
This model, however, overpredicts the resonance lines Fe~{\sc xxv}~w and Fe~{\sc xxvi} Ly$\alpha$ relative to the Fe~{\sc xxv} intercombination and forbidden lines x, y, and z, indicating that the simple optically thin model is insufficient for the detailed Fe K-shell line ratios.
Because the present analysis focuses on the line centroids and their orbital modulation rather than on diagnosing these line ratios, we do not attempt to identify the dominant line-formation process in this narrow-band fit.
Instead, we remove Fe~{\sc xxv}~w and Fe~{\sc xxvi} Ly$\alpha_{1,2}$ from the \texttt{bvcempow} component and model them with Gaussian components.
Each Gaussian is implemented as a \texttt{zgauss} component with a fixed rest-frame transition energy \citep{yerokhin2015,yerokhin2019}.
The redshift, width, and normalization of Fe~{\sc xxv}~w are allowed to vary independently.
For Fe~{\sc xxvi} Ly$\alpha_1$ and Ly$\alpha_2$, we link the redshift and width of the two components and allow their normalizations to vary independently, because the photon statistics are insufficient to constrain two independent Ly$\alpha$ centroids.
The best-fit models and residuals are shown in Figure~\ref{fig:phase_resolved_spectra}, and the best-fit parameters are summarized in Table~\ref{tab:phase_resolved_bestfit}.

We convert the fitted redshifts to line-of-sight velocities and model their orbital modulation as
\[
v(\phi_{\mathrm{orb}}) = \gamma + K_1 \sin\left[2\pi\left(\phi_{\mathrm{orb}} - \phi_0\right)\right],
\]
where $\gamma$ is the mean velocity, $K_1$ is the velocity semi-amplitude, and $\phi_0$ is the phase offset.
Because the fitted line centroids are not converted to the binary rest frame, we treat $\gamma$ as an effective mean velocity rather than as the binary systemic velocity.
It can include gravitational redshift, bulk motion in the accretion column, and observer-frame Doppler offsets.
These terms affect the interpretation of the absolute line centroids, whereas $K_1$ is measured from the orbital-phase-dependent centroid modulation.
We checked uneven sampling of the spacecraft velocity and WD spin phase and found that these effects are not expected to account for the measured modulation (Appendix~\ref{app:phase_systematics}).
The statistical uncertainties of the phase-binned velocities are accounted for in all fits below.
We first evaluate which tracers show orbital modulation by fitting each component independently with either a constant velocity or the sinusoidal model.
The improvement $\Delta\chi^2$ is measured relative to the corresponding constant-velocity fit, with the sinusoidal model adding $K_1$ and $\phi_0$.
The resulting values are summarized in Table~\ref{tab:orbital_modulation_bestfit}.
The strongest improvements are obtained for the Fe~{\sc xxv} tracers: $\Delta\chi^2 = 17.3$ for \texttt{bvcempow}, representing the Fe~{\sc xxv} x, y, and z complex, and 31.8 for \texttt{zgauss 1}, the Fe~{\sc xxv}~w line.
The \texttt{zfeklor} fluorescence component shows a weaker but compatible modulation with $\Delta\chi^2 = 8.5$, whereas the linked \texttt{zgauss 2} + \texttt{zgauss 3} component for Fe~{\sc xxvi} Ly$\alpha_{1,2}$ gives only $\Delta\chi^2 = 2.6$ and does not by itself require orbital modulation.
Because these weaker constraints are driven by larger line-centroid uncertainties, we do not use them to define the representative velocity amplitude.
We therefore adopt the simultaneous fit to the two Fe~{\sc xxv} tracers (i.e., \texttt{bvcempow} and \texttt{zgauss 1}) with shared $K_1$ and $\phi_0$ as the representative measurement of the WD orbital motion, obtaining $K_1 = 58.1 \pm 8.5$~km~s$^{-1}$ and $\phi_0 = 0.51 \pm 0.02$.
The corresponding formal probabilities are listed in Table~\ref{tab:orbital_modulation_bestfit}; for the Fe~{\sc xxv} shared fit, $p \simeq 6 \times 10^{-11}$.
As a consistency check, a simultaneous fit with shared $K_1$ and $\phi_0$ across all four tracers gives $K_1 = 53.9 \pm 7.3$~km~s$^{-1}$, consistent with the adopted Fe~{\sc xxv}-based value.

\begin{figure}[!t]
    \centering
    \includegraphics[width=1.0\linewidth]{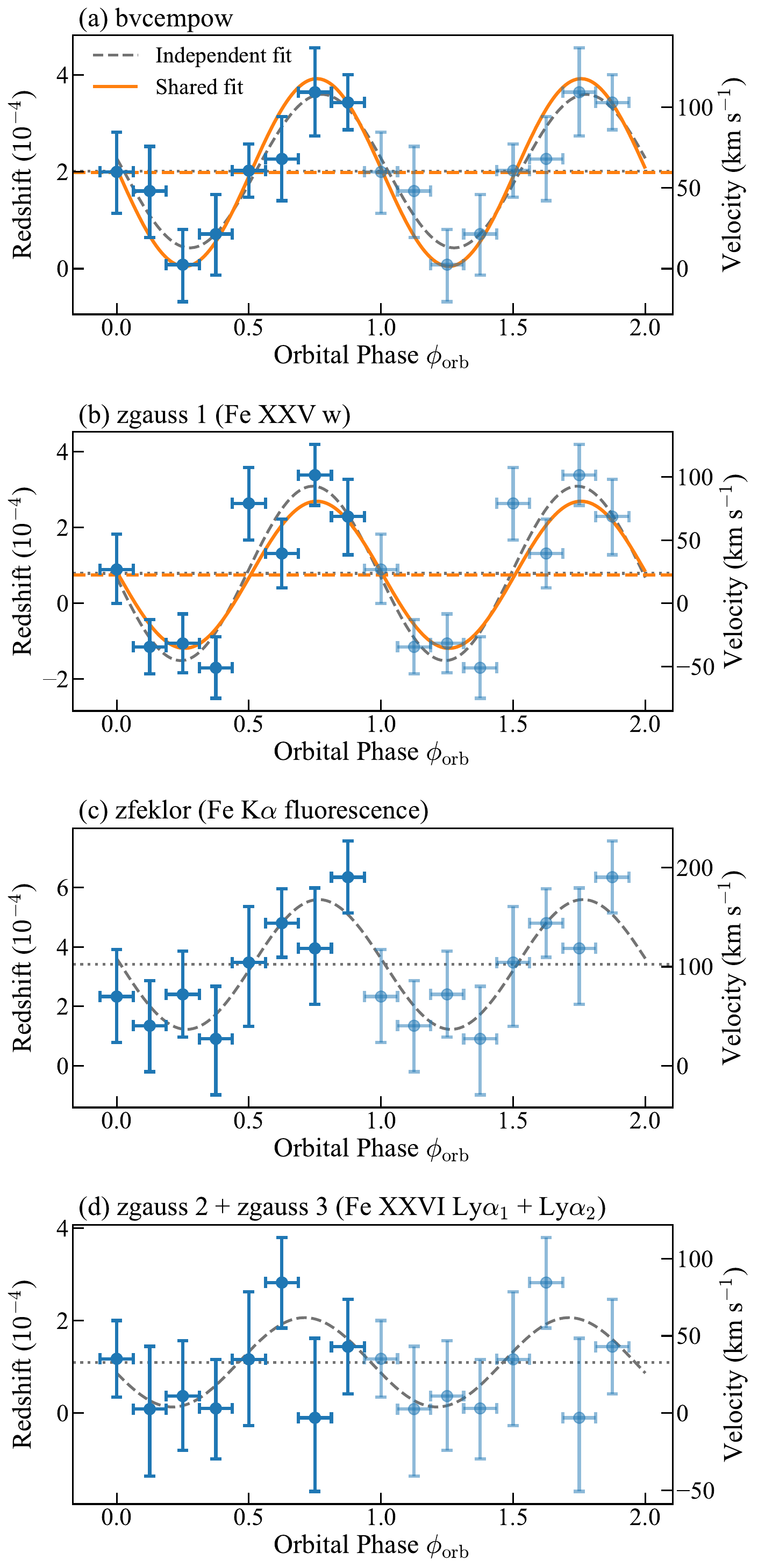}
    \vspace{-4pt}
    \caption
    {
        Orbital modulation of the four spectral components used in the velocity-modulation fit, shown over two cycles by repeating the phase-binned velocities.
        Gray dashed curves and dotted horizontal lines show independent sinusoidal fits and their fitted mean velocities for each component.
        Orange solid curves and dashed horizontal lines are shown only for the two Fe~{\sc xxv} tracers, \texttt{bvcempow} and \texttt{zgauss 1}, and represent the adopted simultaneous fit with shared $K_1$ and $\phi_0$.
        The fit parameters are listed in Table~\ref{tab:orbital_modulation_bestfit}.
    }
    \label{fig:orbital_modulation_summary}
\end{figure}

\floattable
\begin{deluxetable}{llcccccccc}
\tablewidth{0pt}
\tablecaption{Best-fit parameters for each phase bin.}
\label{tab:phase_resolved_bestfit}
\tablehead{
    \multicolumn{1}{l}{Component}
    & \multicolumn{1}{l}{Parameter (Unit)}
    & \multicolumn{1}{c}{Phase 0}
    & \multicolumn{1}{c}{Phase 1}
    & \multicolumn{1}{c}{Phase 2}
    & \multicolumn{1}{c}{Phase 3}
    & \multicolumn{1}{c}{Phase 4}
    & \multicolumn{1}{c}{Phase 5}
    & \multicolumn{1}{c}{Phase 6}
    & \multicolumn{1}{c}{Phase 7}
}
\startdata
\texttt{zfeklor} & Redshift ($10^{-4}$)
& $2.3_{-1.5}^{+1.6}$
& $1.3_{-1.5}^{+1.5}$
& $2.4_{-1.4}^{+1.5}$
& $0.9_{-1.9}^{+1.8}$
& $3.5_{-2.2}^{+1.9}$
& $4.8_{-1.2}^{+1.2}$
& $4.0_{-1.9}^{+2.0}$
& $6.3_{-1.2}^{+1.2}$ \\
(Fe~K$\alpha$ fluorescence) & Normalization\tablenotemark{a}
& $2.1_{-0.4}^{+0.5}$
& $3.5_{-0.6}^{+0.6}$
& $2.6_{-0.5}^{+0.6}$
& $1.9_{-0.5}^{+0.5}$
& $2.1_{-0.5}^{+0.6}$
& $4.6_{-0.8}^{+0.9}$
& $2.0_{-0.5}^{+0.5}$
& $2.4_{-0.5}^{+0.5}$ \\
\texttt{bvcempow} & $T_{\max}$ (keV)
& $11.9_{-0.8}^{+0.9}$
& $11.6_{-0.9}^{+0.9}$
& $12.3_{-0.8}^{+1.0}$
& $10.0_{-0.7}^{+0.8}$
& $11.7_{-1.0}^{+1.0}$
& $11.0_{-0.9}^{+1.0}$
& $11.7_{-0.8}^{+1.0}$
& $11.3_{-0.8}^{+0.8}$ \\
& Redshift ($10^{-4}$)
& $2.0_{-0.9}^{+0.8}$
& $1.6_{-1.0}^{+0.9}$
& $0.1_{-0.8}^{+0.7}$
& $0.7_{-0.8}^{+0.8}$
& $2.0_{-0.6}^{+0.5}$
& $2.3_{-0.9}^{+0.9}$
& $3.6_{-0.9}^{+0.9}$
& $3.4_{-0.6}^{+0.6}$ \\
& $\sigma_v$ (km s$^{-1}$)\tablenotemark{b}
& $131_{-26}^{+27}$
& $139_{-34}^{+39}$
& $102_{-35}^{+33}$
& $141_{-28}^{+31}$
& $<77$
& $118_{-30}^{+33}$
& $150_{-30}^{+34}$
& $71_{-43}^{+32}$ \\
& Normalization\tablenotemark{c}
& $0.160_{-0.007}^{+0.007}$
& $0.168_{-0.008}^{+0.009}$
& $0.172_{-0.008}^{+0.008}$
& $0.190_{-0.010}^{+0.010}$
& $0.170_{-0.008}^{+0.010}$
& $0.198_{-0.011}^{+0.013}$
& $0.178_{-0.009}^{+0.009}$
& $0.181_{-0.008}^{+0.009}$ \\
\texttt{zgauss 1}
& Energy (eV)
& \multicolumn{8}{c}{6700.43 (fixed)} \\
(Fe~{\sc xxv} w) & Sigma (eV)
& $6.2_{-0.6}^{+0.6}$
& $4.6_{-0.5}^{+0.5}$
& $5.3_{-0.5}^{+0.6}$
& $5.0_{-0.5}^{+0.5}$
& $5.7_{-0.7}^{+0.7}$
& $4.8_{-0.5}^{+0.6}$
& $4.7_{-0.6}^{+0.7}$
& $6.2_{-0.7}^{+0.7}$ \\
& Redshift ($10^{-4}$)
& $0.9_{-0.9}^{+0.9}$
& $-1.2_{-0.7}^{+0.7}$
& $-1.1_{-0.8}^{+0.8}$
& $-1.7_{-0.8}^{+0.8}$
& $2.6_{-1.0}^{+0.9}$
& $1.3_{-0.9}^{+0.9}$
& $3.4_{-0.8}^{+0.8}$
& $2.3_{-1.0}^{+1.0}$ \\
& Normalization\tablenotemark{a}
& $11.2_{-0.9}^{+1.0}$
& $11.7_{-1.0}^{+1.0}$
& $12.4_{-1.0}^{+1.1}$
& $10.8_{-0.9}^{+1.0}$
& $11.4_{-1.1}^{+1.1}$
& $12.0_{-1.2}^{+1.3}$
& $10.4_{-1.0}^{+1.1}$
& $10.4_{-0.9}^{+1.0}$ \\
\texttt{zgauss 2}
& Energy (eV)
& \multicolumn{8}{c}{6973.18 (fixed)} \\
(Fe~{\sc xxvi} Ly$\alpha_1$) & Sigma (eV)
& $4.7_{-0.5}^{+0.5}$
& $6.1_{-0.9}^{+1.3}$
& $6.1_{-0.6}^{+0.7}$
& $5.8_{-0.6}^{+0.8}$
& $6.1_{-0.7}^{+0.9}$
& $4.8_{-0.5}^{+0.6}$
& $6.4_{-0.8}^{+1.1}$
& $5.4_{-0.6}^{+0.7}$ \\
& Redshift ($10^{-4}$)
& $1.2_{-0.8}^{+0.8}$
& $0.1_{-1.4}^{+1.4}$
& $0.4_{-1.2}^{+1.2}$
& $0.1_{-1.1}^{+1.1}$
& $1.2_{-1.4}^{+1.5}$
& $2.8_{-1.0}^{+1.0}$
& $-0.1_{-1.6}^{+1.7}$
& $1.4_{-1.0}^{+1.0}$ \\
& Normalization\tablenotemark{a}
& $5.6_{-0.6}^{+0.7}$
& $5.5_{-0.7}^{+0.8}$
& $7.7_{-0.8}^{+0.9}$
& $6.3_{-0.7}^{+0.8}$
& $5.2_{-0.8}^{+0.8}$
& $6.9_{-0.9}^{+1.0}$
& $5.9_{-0.8}^{+0.9}$
& $6.2_{-0.7}^{+0.8}$ \\
\texttt{zgauss 3}
& Energy (eV)
& \multicolumn{8}{c}{6951.97 (fixed)} \\
(Fe~{\sc xxvi} Ly$\alpha_2$) & Sigma (eV)
& \multicolumn{8}{c}{linked to Fe~{\sc xxvi} Ly$\alpha_1$} \\
& Redshift ($10^{-4}$)
& \multicolumn{8}{c}{linked to Fe~{\sc xxvi} Ly$\alpha_1$} \\
& Normalization\tablenotemark{a}
& $3.9_{-0.5}^{+0.6}$
& $3.5_{-0.6}^{+0.7}$
& $4.5_{-0.7}^{+0.7}$
& $4.6_{-0.6}^{+0.7}$
& $5.7_{-0.8}^{+0.8}$
& $5.7_{-0.9}^{+0.9}$
& $4.4_{-0.7}^{+0.8}$
& $4.5_{-0.6}^{+0.7}$ \\ \hline
Cstat &
& 2227.8
& 2231.8
& 2214.1
& 2196.2
& 2080.5
& 1987.0
& 2203.7
& 2276.4 \\
d.o.f. &
& 1986
& 1986
& 1986
& 1986
& 1986
& 1986
& 1986
& 1986 \\
\enddata
\tablenotetext{a}{$10^{-5}$ photons cm$^{-2}$ s$^{-1}$}
\tablenotetext{b}{Velocity dispersion in the line of sight for the Doppler broadening of the emission lines.}
\tablenotetext{c}{Normalization is defined as $\frac{10^{-14}}{4 \pi D^2} \int n_{\mathrm{e}} n_{\mathrm{H}} \mathrm{d}V$, where $D$ is the distance to the source, $n_{\mathrm{e}}$ is the electron number density, and $n_{\mathrm{H}}$ is the hydrogen number density.}
\tablecomments{
For \texttt{bvcempow}, the unlisted fixed parameters are $\alpha = 0.5$, $n_{\mathrm{H}} = 1~\mathrm{cm}^{-3}$, elemental abundances fixed to solar values, and switch = 2; the last setting computes the spectrum based on the AtomDB database.
The atomic-data references follow the compilation of \citet{hell2025}.
The fixed \texttt{zfeklor} profile uses the laboratory Fe~{\sc i} K$\alpha$ energies and relative intensities of \citet{holzer1997}.
The fixed rest-frame energies of Fe~{\sc xxv} w and Fe~{\sc xxvi} Ly$\alpha_{1,2}$ are from \citet{yerokhin2015,yerokhin2019}.
}
\end{deluxetable}

\begin{deluxetable}{lcccccc}
\tablewidth{0pt}
\tablecaption{Velocity-modulation fits to the Fe K-shell line centroids.}
\label{tab:orbital_modulation_bestfit}
\tablehead{
    \colhead{Component}
    & \colhead{$\gamma$ (km s$^{-1}$)}
    & \colhead{$K_1$ (km s$^{-1}$)}
    & \colhead{$\phi_0$}
    & \colhead{$\chi^2/\mathrm{dof}$}
    & \colhead{$\Delta\chi^2$}
    & \colhead{Formal $p$}
}
\startdata
\multicolumn{7}{c}{Independent sinusoidal fits to each tracer} \\ \hline
\texttt{zfeklor}
& $102.3 \pm 16.1$
& $65.5 \pm 22.6$
& $0.51 \pm 0.06$
& $3.5/5$
& 8.5
& $1.4 \times 10^{-2}$ \\
\texttt{bvcempow}
& $60.4 \pm 7.9$
& $47.5 \pm 11.6$
& $0.53 \pm 0.04$
& $1.5/5$
& 17.3
& $1.7 \times 10^{-4}$ \\
\texttt{zgauss 1}
& $23.7 \pm 9.0$
& $69.0 \pm 12.2$
& $0.49 \pm 0.03$
& $7.0/5$
& 31.8
& $1.2 \times 10^{-7}$ \\
\texttt{zgauss 2} + \texttt{zgauss 3}
& $32.9 \pm 12.0$
& $28.9 \pm 18.1$
& $0.46 \pm 0.09$
& $2.9/5$
& 2.6
& 0.27 \\ \hline
\multicolumn{7}{c}{Simultaneous fits with shared $K_1$ and $\phi_0$} \\ \hline
\multicolumn{7}{c}{Fe~{\sc xxv} shared fit} \\ \hline
\texttt{bvcempow}
& $59.6 \pm 7.8$
& \multirow[c]{2}{*}{$58.1 \pm 8.5$}
& \multirow[c]{2}{*}{$0.51 \pm 0.02$}
& \multirow[c]{2}{*}{$10.6/12$}
& \multirow[c]{2}{*}{47.0}
& \multirow[c]{2}{*}{$6.2 \times 10^{-11}$} \\
\texttt{zgauss 1}
& $22.6 \pm 9.0$
&
&
&
&
&  \\ \hline
\multicolumn{7}{c}{All-component shared fit} \\ \hline
\texttt{zfeklor}
& $103.5 \pm 15.9$
& \multirow[c]{4}{*}{$53.9 \pm 7.3$}
& \multirow[c]{4}{*}{$0.50 \pm 0.02$}
& \multirow[c]{4}{*}{$19.9/26$}
& \multirow[c]{4}{*}{55.1}
& \multirow[c]{4}{*}{$1.1 \times 10^{-12}$} \\
\texttt{bvcempow}
& $59.7 \pm 7.8$
&
&
&
&
&  \\
\texttt{zgauss 1}
& $22.3 \pm 9.0$
&
&
&
&
&  \\
\texttt{zgauss 2} + \texttt{zgauss 3}
& $31.7 \pm 11.9$
&
&
&
&
&  \\ \hline
\enddata
\tablecomments{
$\Delta\chi^2$ is measured relative to the corresponding constant-velocity fit.
For simultaneous fits, the constant-velocity comparison allows each component to have its own mean velocity, while the sinusoidal model adds a shared $K_1$ and $\phi_0$.
Formal $p$ values are computed by treating $\Delta\chi^2$ as a likelihood-ratio statistic with two additional sinusoidal parameters, i.e., using the $\chi^2_2$ approximation.
They do not include possible systematic uncertainties in the line-centroid measurements.
}
\end{deluxetable}

\section{Discussion} \label{sec: discussion}

Enabled by the high spectral resolution and in-orbit gain calibration of Resolve, we detected coherent orbital modulation in the centroids of the Fe~{\sc xxv} K$\alpha$ components in EX~Hya (Figure~\ref{fig:orbital_modulation_summary}).
The independently fitted amplitudes and phases of the two Fe~{\sc xxv} tracers, \texttt{bvcempow} and \texttt{zgauss 1}, are mutually consistent, supporting the interpretation that their centroid modulation follows the center-of-mass orbital motion of the accreting WD.
We therefore adopt the simultaneous fit to these Fe~{\sc xxv} tracers as the representative measurement, $K_1 = 58.1 \pm 8.5$~km~s$^{-1}$ (Table~\ref{tab:orbital_modulation_bestfit}).
We do not use the Fe K$\alpha$ fluorescence and linked Fe~{\sc xxvi} Ly$\alpha_{1,2}$ components to define the representative $K_1$, because their larger centroid uncertainties make their individual modulation tests less constraining.
Including them in the shared-modulation fit provides a consistency check, giving $K_1 = 53.9 \pm 7.3$~km~s$^{-1}$, consistent with the adopted Fe~{\sc xxv}-based value.
This X-ray velocity amplitude is consistent with optical and ultraviolet emission-line measurements near 58~km~s$^{-1}$ \citep{belle2003,echevarria2016,beuermann2024} and with the previous Chandra/HETG X-ray measurement of $58.2 \pm 3.7$~km~s$^{-1}$ \citep{hoogerwerf2004}.
The key advance over previous X-ray radial-velocity work is that the WD velocity amplitude is recovered from individual Fe K-shell line modulation rather than from a many-line average.
The agreement with the optical/UV values is important because line centroids formed in physically different regions nevertheless recover the same WD orbital motion.

The measured $K_1$ allows us to derive the WD mass when combined with binary parameters from previous dynamical studies.
For a circular binary orbit, the mass function associated with the secondary-star velocity amplitude $K_2$ is
\begin{equation} \label{eq:mass_function}
f_1 \equiv \frac{(M_1 \sin i)^3}{(M_1 + M_2)^2} = \frac{P K_2^3}{2\pi G},
\end{equation}
where $M_1$ and $M_2$ are the masses of the primary and secondary stars, respectively, $i$ is the orbital inclination, $P$ is the orbital period, and $G$ is the gravitational constant.
Combining Equation~\ref{eq:mass_function} with the velocity-ratio relation $K_1/K_2 = M_2/M_1$, we obtain
\begin{equation}  \label{eq:M1}
M_1 = \left(\frac{P}{2\pi G}\right) \frac{K_2^3 (1+K_1/K_2)^2}{\sin^3 i}.
\end{equation}
Adopting $P = 0.068233843$~d from \citet{echevarria2016} and $K_2 = 432.4 \pm 4.8$~km~s$^{-1}$ and $i = 77.8^\circ \pm 0.4^\circ$ from \citet{beuermann2008}, together with the Fe~{\sc xxv}-based $K_1 = 58.1 \pm 8.5$~km~s$^{-1}$, we obtain $M_1 = 0.79 \pm 0.04~M_\odot$.
Using instead the all-component shared-fit value, $K_1 = 53.9 \pm 7.3$~km~s$^{-1}$, gives $M_1 = 0.77 \pm 0.03~M_\odot$, consistent with the adopted value.
The mass changes little between these two choices because $K_1$ enters Equation~\ref{eq:M1} only through the small correction factor $(1+K_1/K_2)^2$, with $K_1 \ll K_2$; the error propagation is summarized in Appendix~\ref{app:mass_error}.
This estimate still uses the literature $K_2$ and inclination, but its new ingredient is the X-ray measurement of the WD velocity amplitude from Fe~{\sc xxv} line centroids.

The adopted dynamical mass is consistent with previous optical/UV dynamical estimates, as expected from the agreement between the X-ray and optical/UV $K_1$ values.
A more physically distinct comparison is with X-ray spectral modeling.
\citet{hayashi2014a} obtained $M_1 = 0.63^{+0.17}_{-0.14}~M_\odot$ from Suzaku post-shock accretion-column modeling, whereas \citet{suleimanov2019} obtained $M_1 = 0.70 \pm 0.04~M_\odot$ from NuSTAR hard-X-ray spectral modeling.
These spectral-model masses are somewhat lower than, but not in strong tension with, our dynamical estimate, illustrating the sensitivity of X-ray spectroscopic mass estimates to accretion-column geometry and shock-height assumptions.
Thus, EX~Hya provides a useful benchmark for comparing dynamical X-ray Doppler measurements with X-ray spectral-model mass estimates.

The main caveat is that the component-dependent mean velocities require line-by-line interpretation.
In the simultaneous fit, the Fe K$\alpha$ fluorescence component has the largest mean velocity, $\gamma = 103.5 \pm 15.9$~km~s$^{-1}$, whereas the highly ionized components give lower values of $\gamma \simeq 20$--$60$~km~s$^{-1}$ (Table~\ref{tab:orbital_modulation_bestfit}).
This ordering is physically plausible: Fe K$\alpha$ fluorescence is expected to arise from low-ionization material on or near the WD surface \citep[e.g.,][]{hayashi2018}, where gravitational redshift is largest, whereas the highly ionized lines are produced in hotter plasma at larger heights in the post-shock flow \citep[e.g.,][]{hayashi2014}.
The absolute mean velocity of the fluorescence component also depends on the ionization state; a separate analysis of the Fe K$\alpha$ and K$\beta$ centroids favors Fe~{\sc vi} rather than neutral Fe (Terada et al. in prep.).
At the same time, the simple optically thin plasma model overpredicts the resonance lines Fe~{\sc xxv}~w and Fe~{\sc xxvi} Ly$\alpha$ relative to the Fe~{\sc xxv} intercombination and forbidden lines.
Resonance scattering is a plausible explanation for this behavior, especially because such effects have been predicted \citep{terada2001} and recently detected in the accretion column in AM~Her \citep{terada2026}.
Identifying the dominant line-formation process would require dedicated radiative-transfer modeling, which is beyond the scope of the present orbital-modulation analysis.
This uncertainty mainly affects the interpretation of the line ratios and component-dependent mean velocities; it does not change the Fe~{\sc xxv}-based measurement of $K_1$, which is determined by the phase-dependent centroid modulation.

\section{Conclusions} \label{sec: Conclusions}

We analyzed the 83~ks XRISM/Resolve observation of the intermediate polar EX~Hya to measure the orbital modulation of Fe K-shell line centroids.
Phase-resolved spectroscopy reveals coherent orbital modulation in the Fe~{\sc xxv} K$\alpha$ components.
This is the first detection of orbital modulation in individual Fe K-shell lines from an accreting WD, rather than in a many-line average.
A simultaneous fit to the two Fe~{\sc xxv} tracers gives $K_1 = 58.1 \pm 8.5$~km~s$^{-1}$.
The Fe K$\alpha$ fluorescence and Fe~{\sc xxvi} Ly$\alpha_{1,2}$ components are less constraining individually, but remain consistent with the Fe~{\sc xxv}-based modulation.
The consistency between the X-ray and optical/UV $K_1$ measurements provides an important cross-check that both velocity tracers follow the WD orbital motion, because they arise from physically distinct line-forming regions.
Combining this X-ray measurement with literature values of $K_2$ and the inclination yields a WD mass of $M_1 = 0.79 \pm 0.04~M_\odot$.
Thus, high-resolution X-ray spectroscopy with Resolve establishes the centroid modulation of individual Fe K-shell lines as a complementary X-ray method for measuring $K_1$ in accreting WDs.

\begin{acknowledgments}
The author thanks the XRISM team for the operation of the observatory and the preparation of the calibration products used in this work.
This work was supported by JSPS KAKENHI Grant Numbers JP26K17204 (Y.O.), JP20K04009 (Y.T.), JP21K13970 (M.K.), and JP24H01807 (M.K.), and by the JSPS Core-to-Core Program under Grant Number JPJSCCA20220002.
Y.O. acknowledges support from the Special Postdoctoral Researchers Program in RIKEN.
\end{acknowledgments}

\facilities{XRISM (Resolve)}

\software{HEASoft, XSPEC}

\appendix 

\section{Checks on phase-dependent systematics} \label{app:phase_systematics}

\subsection{\texorpdfstring{Sampling of the projected spacecraft velocity}{Sampling of the projected spacecraft velocity}}

\begin{figure*}[tbp]
    \centering
    \includegraphics[width=1.0\textwidth]{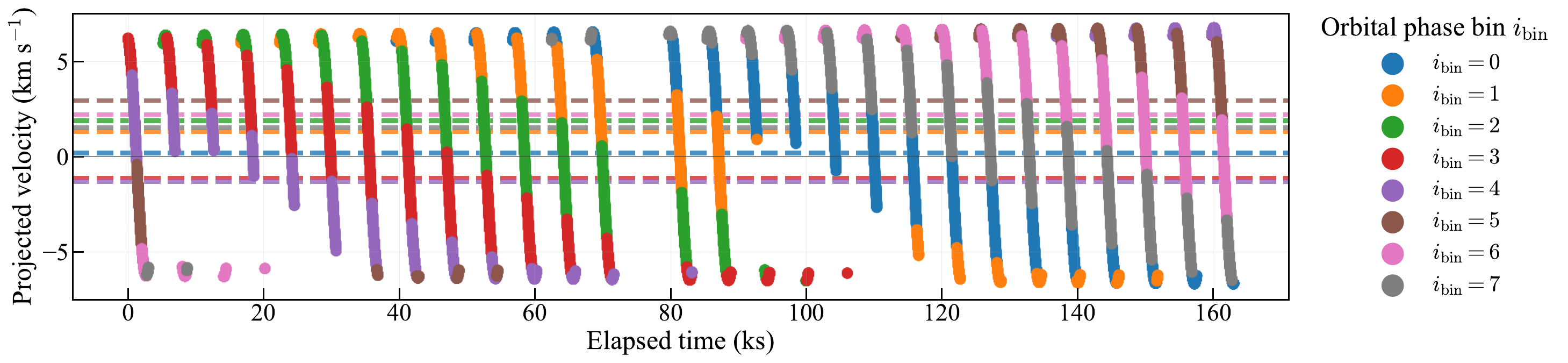}
    \vspace{-4pt}
    \caption{
        Line-of-sight component of the XRISM spacecraft velocity toward EX~Hya evaluated for screened events.
        Positive values correspond to spacecraft motion toward EX~Hya.
        Points are colored by the orbital phase bins used for the phase-resolved spectroscopy, and horizontal dashed lines show the phase-bin averages.
    }
    \label{fig:orbital_velocity_projection_by_phase}
\end{figure*}

Because the phase-resolved spectra are accumulated from non-continuous screened exposure intervals, uneven sampling of XRISM's Earth-orbital velocity projected along the line of sight to EX~Hya could in principle introduce a phase-dependent shift in the measured line centroids.
To check this effect, we assigned each screened event the same EX~Hya orbital phase and phase bin as used for the phase-resolved spectroscopy in Section~\ref{sec: method}.
We then interpolated the XRISM spacecraft velocity taken from the orbit file to each event time and projected it onto the line of sight toward EX~Hya.
This event-based calculation includes the non-continuous exposure and the actual distribution of detected photons by construction.
Figure~\ref{fig:orbital_velocity_projection_by_phase} shows the resulting line-of-sight velocity component.
The maximum difference among the phase-bin mean velocities is $\sim 4$~km~s$^{-1}$, smaller than the statistical uncertainty on the adopted Fe~{\sc xxv}-based $K_1 = 58.1 \pm 8.5$~km~s$^{-1}$.
Thus, uneven sampling of the projected spacecraft velocity is not expected to account for the measured orbital modulation.

\subsection{\texorpdfstring{Sampling of the WD spin phase}{Sampling of the WD spin phase}}

\begin{figure*}[tbp]
    \centering
    \includegraphics[width=1.0\textwidth]{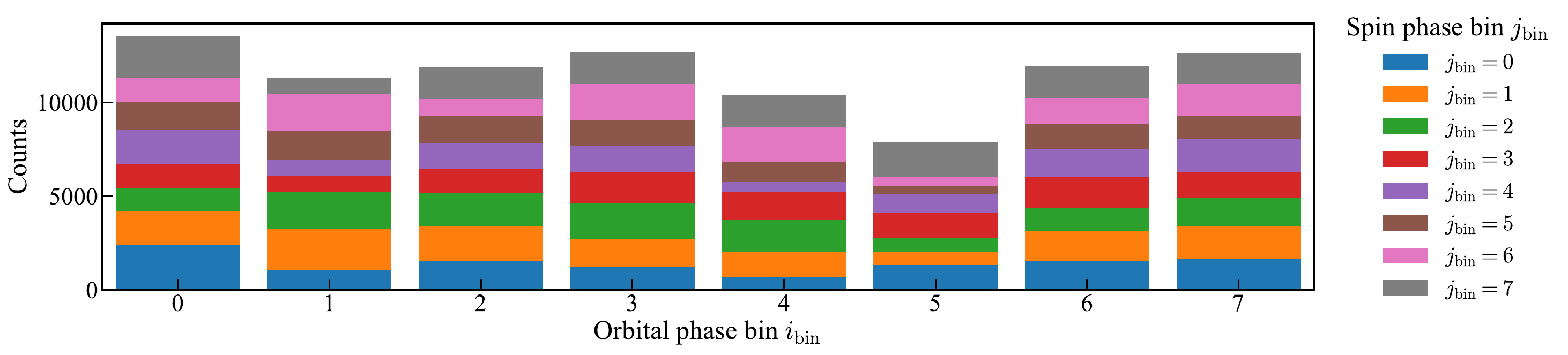}
    \vspace{-4pt}
    \caption{
        Spin-phase sampling within the orbital phase bins for the 1.7--10~keV screened events.
        The bar height gives the event counts in each orbital phase bin $i_{\mathrm{bin}}$, and the stacked colored segments show the contributions from the WD spin-phase bins $j_{\mathrm{bin}}$.
        Both sets of bins use the centered-bin definition adopted for the phase-resolved spectroscopy.
        The orbital phase is calculated with the eclipse ephemeris of \citet{echevarria2016}, and the WD spin phase with the quadratic spin-pulse ephemeris of \citet{beuermann2024a}.
    }
    \label{fig:phase_coverage_orb_spin}
\end{figure*}

Uneven sampling of the WD spin phase within the binary orbital bins could in principle bias the phase-resolved line measurements.
This is a relevant check because EX~Hya shows a well-known 67~min WD spin modulation.
In this high-inclination system, the accretion column is viewed largely side-on, so the spin modulation is expected to affect the observed flux and spectral shape mainly through occultation or absorption in the accretion-column or accretion-curtain geometry rather than through large line-of-sight velocity shifts \citep{rosen1988,allan1998,pekon2011}.
To check this effect, we computed the WD spin phase of each screened event using the quadratic spin-pulse ephemeris of \citet{beuermann2024a}, which gives the BJD(TDB) times of the 67~min spin-pulse maxima.
The orbital phase and orbital phase bin were computed with the same eclipse ephemeris of \citet{echevarria2016} and the same centered-bin definition as used for the phase-resolved spectroscopy.
Figure~\ref{fig:phase_coverage_orb_spin} shows the 1.7--10~keV event counts in each orbital phase bin, with the contributions from the WD spin-phase bins shown as stacked components.
Although the spin-phase coverage is not uniform, no orbital phase bin is dominated by a single spin-phase bin.
Thus, uneven spin-phase sampling is not expected to account for the measured orbital modulation.

\section{Error propagation for the dynamical mass} \label{app:mass_error}

We estimate the uncertainty in the dynamical mass by propagating the
independent statistical errors in $K_1$, $K_2$, and $i$ through
Equation~\ref{eq:M1}.
Taking the logarithmic differential gives
\begin{equation}
\frac{\mathrm{d}M_1}{M_1}
=
\frac{2\,\mathrm{d}K_1}{K_1+K_2}
+
\left(\frac{1}{K_2}+\frac{2}{K_1+K_2}\right)\mathrm{d}K_2
-
3\cot i\,\mathrm{d}i .
\end{equation}
Assuming independent errors and neglecting the uncertainty in $P$, the fractional uncertainty is
\begin{equation}
\left(\frac{\sigma_{M_1}}{M_1}\right)^2 =
\left[\frac{2\sigma_{K_1}}{K_1+K_2}\right]^2
+ \left[\left(\frac{1}{K_2}+\frac{2}{K_1+K_2}\right)\sigma_{K_2}\right]^2
+ \left(3\cot i\,\sigma_i\right)^2,
\end{equation}
where $\sigma_{K_1}$, $\sigma_{K_2}$, and $\sigma_i$ are the 1$\sigma$
uncertainties in $K_1$, $K_2$, and $i$, respectively, with $\sigma_i$
expressed in radians.
Using the Fe~{\sc xxv}-based $K_1$ value adopted in Section~\ref{sec: discussion}, this gives a total fractional mass uncertainty of $\sigma_{M_1}/M_1 = 4.7\%$.
For $M_1 = 0.79~M_\odot$, this corresponds to $\sigma_{M_1} \simeq 0.037~M_\odot$, consistent with the quoted uncertainty of $0.04~M_\odot$.
If only one input uncertainty is included at a time, the corresponding values of $\sigma_{M_1}/M_1$ are 3.5\%, 3.1\%, and 0.45\% for the uncertainties in $K_1$, $K_2$, and $i$, respectively.
The $K_1$ term is much smaller than the fractional uncertainty in $K_1$ itself because $K_1$ enters Equation~\ref{eq:M1} only through the factor $(1+K_1/K_2)^2$, or equivalently through the sum $K_1+K_2$ in the differential above, with $K_1 \ll K_2$.

\bibliography{exhya_orbital_modulation}{}
\bibliographystyle{aasjournalv7}

\end{document}